\documentclass[12pt]{iopart}
\usepackage{graphicx}
\usepackage{dcolumn}
\usepackage{bm}
\usepackage{hyperref}
\usepackage[mathlines]{lineno}
\usepackage{subfigure}

\newcommand{\bra}[1]{\left\langle #1\right|}
\newcommand{\ket}[1]{\left|#1\right\rangle}

\usepackage{iopams}
\begin{document}

\title{Survival probability in a one-dimensional quantum walk on a trapped lattice}
\author{Meltem G\"{o}n\"{u}lol}
\address{Department of Physics, Dokuz
Eyl\"{u}l University, Tr-35160 \.{I}zmir, Turkey}
\ead{g.meltem@gmail.com}
\author{Ekrem Ayd{\i}ner}
 \address{Department of Physics, \.{I}stanbul University, Tr-34134,
\.{I}stanbul, Turkey} \ead{ekrem.aydiner@istanbul.edu.tr}
\author{Yutaka Shikano}
 \address{Department of Physics,
Tokyo Institute of Technology, Tokyo 152-8551, Japan}
\address{Department of Mechanical Engineering, Massachusetts Institute of
Technology, Cambridge, MA 02139,USA}
\ead{shikano@th.phys.titech.ac.jp}
\author{\"{O}zg\"{u}r E. M\"{u}stecapl{\i}o\~{g}lu}
 \address{Department of Physics, Ko\c{c} University, \.{I}stanbul, 34450,
Turkey} \ead{omustecap@ku.edu.tr}

\begin{abstract}
The dynamics of the survival probability of quantum walkers on a one-dimensional lattice with random distribution of absorbing immobile
traps are investigated. The survival probability of quantum walkers is compared with that of classical
walkers. It is shown that the time dependence of survival
probability of quantum walkers has a piecewise stretched exponential
character depending on the density of traps  in numerical and analytical observations.
The crossover between the quantum analogs of the Rosenstock and Donsker-Varadhan behaviors
is identified.
\end{abstract}

\maketitle

\section{\label{sec:intro}Introduction}
The classical random walk (CRW) is a prototype model of stochastic
processes that occur in many physical systems \cite{barber}. Extension of
random walk concept from stochastic classical realm to the unitarily
evolving quantum world is motivated by the promise of quantum walks
(QWs) \cite{ambainis2001,kempe} as quantum algorithms \cite{shenvi}
outperforming their classical counterparts, and as a simple model for
quantum computation.

In parallel with the remarkable developments in the experimental
ability to control single atoms and photons, early proposals and
demonstrations of QWs are followed by more robust and controllable
implementations \cite{zahringer,schmitz,schreiber}.
Superiority of a QW algorithm \cite{shenvi} is experimentally
demonstrated very recently \cite{lu}. In a latest experiment
\cite{broome}, the effect of absorbing boundaries on the quantum
walk is examined. The QW on a line segment with absorbing boundaries
\cite{ambainis2001,Konno2} is a special case of a more general
situation of QW in the presence of absorbing traps.

A QW on a trapped lattice exhibits transition to CRW \cite{Gonulol};
therefore traps can be characterized as a quantum decoherence
mechanism \cite{kendon2006, CKSS1}, similar to broken links
\cite{romanelli} or environmental noise \cite{chandra2007}.
Controlling trap density in the lattice allows for tunable
decoherence mechanism, which is beneficial for fundamental
investigations of quantum decoherence and for efficient
implementation and speeding up quantum algorithms
\cite{kendon,Whitfield}.

In addition to their role as a source of quantum decoherence,
traps can play another role on the dynamics of QW.
It is known that their presence causes different dynamical regimes
on the evolution of CRW. Our aim is to explore if such
distinct dynamical regimes can emerge in QW without changing
its quantum nature. Earlier occurrences of such a crossover between
different dynamical regimes than quantum to classical transition
should be taken into account potential applications of trapped QW.

Trapped CRW was extensively explored
\cite{havlin_review,trapCRW1,trapCRW4}. A practical quantity of
interest is the survival probability of diffusing particles, which
is the mean probability that a walker can still be found on the
lattice after some time $t$. It can be analytically
calculated for a one-dimensional CRW \cite{anlauf}. In the early
times of CRW on a one-dimensional lattice with low trap
concentration, survival probability decays exponentially with the
square root of time, $t^{1/2}$, which is known as Rosenstock (RS)
\cite{RS} behavior. At asymptotically large times, this behavior
makes a crossover \cite{Barkema} to a qualitatively different
scaling form, which is called  Donkser and Varadhan (DV) regime
\cite{DV,Havlin1,Grassberger,haus}, in which the survival
probability exponentially decays with $t^{1/3}$. Similar scaling
forms also appear in the closely related problem of Lifshitz tail or
Griffiths singularity of the density of states at the band edge for
a quantum electron in random potential
\cite{lifshitz,Jayannavar3,nieuwenhuizen}.

It is neither intuitively nor quantitatively obvious to extend the
classical results to characterize the survival probability of
quantum walkers on a trapped land, because of the curious role of
quantum coherence and path interference played in a QW which is
associated with the characteristic strong de-localization of quantum
walkers. This paper specifically addresses the question of quantum
diffusion dynamics on a trapped chain, in particular investigates
the quantum analogs of the RS and DV dynamical regimes of the
classical diffusion problem.

It is difficult to observe the RS-to-DV crossover in CRW. Time
dependence of classical survival probability makes the crossover
only after a long time, though it  can happen relatively earlier for
larger trap concentrations or for larger diffusive constants
\cite{schotland}. Motivated by the role of the diffusive constant in
shortening the crossover time in CRW, we predict that highly
de-localized quantum walkers can enter the DV regime earlier than
the classical walkers. Furthermore, the qualitatively different DV
scaling form in CRW is attributed to the existence of large voids,
or absorber-free regions, which are exponentially rare among the
possible configurations \cite{Grassberger}. Their contribution can
dominate the time dependence of the survival probability only at
large times, after the more common smaller voids lost their walkers.
In QW, we expect that, due to the larger spread of the walkers, such
voids should be larger, and hence more rare. An average over such
clusters, with their corresponding large decay times, would lead to
slower decay of quantum walkers than their classical counterparts.
Indeed, for a continuous time one-dimensional quantum transport
problem, it is found that the survival probability exponentially
decays with $t^{1/4}$ \cite{Parris2}. This asymptotically slower
decay of quantum coherent particles than the diffusive classical
particles is explained by the existence of slowly-decaying
asymptotically large trap-free segments \cite{Parris2}. In QW-based
search algorithms with multi-agents, such slowing down of quantum
coherent dynamics would cause an additional limitation of the
quantum speeding up even for small target (or trap) concentrations.
It should be taken into account in addition to the usual
quantum-to-classical decoherence problem.

We perform detailed numerical
simulations for settings relevant to current experimental efforts. As such, our
discussion is limited to the one-dimensional coined discrete-time QW.
Comparative studies of signatures of coherent and incoherent
transport in the case of continuous-time quantum walk are recently reported
 \cite{Parris2,mulken,agliari}. We give particular attention
to the practical case of small, finite size lattices and small
number of time steps. Similar to the classical prototype system of
disordered media, randomly distributed static traps are assumed.
The trapping process is supposed to be a quenched, instantaneous and
perfect absorption of walkers. In a typical scenario of interest
there would be few traps, and dynamics would be limited to short
times; but the cases of long time behavior as well as densely
trapped lattices are also analyzed to comprehend differences in both
RS and DV regimes, in addition to dynamics of quantum-to-classical
transition.

We explain our numerical results by the Flory-type heuristic
arguments \cite{florry} used in polymer chemistry.
Spatial arrangements of macromolecules, or conformations of polymers, are closely related to
the diffusion and the random walk problem. In the early 1930s, structural chemistry descriptions of
long-chain molecules were based upon unconstrained random walks, where the skeletal bonds of
the molecule are represented by the uncorrelated steps of random walkers. This analogy yields
scaling relations for the root mean square (rms) distance of the chain (squared radius of gyration)
depending on the bond length and the number of the bonds. In 1949, P. J. Florry has provided
a seminal work which takes into account volume exclusion effect (no segments of a molecule can
overlap in space), in formation of polymers. This allows for description of polymer growth in terms of
the self-avoiding or repulsive random walks. In self-avoiding walks, the walker would stop or become trapped if there are no
more unvisited neighboring sites. Our trapped lattice model is in that sense is closely related to
such random walk models of polymer growth and size distribution. The movement of a single walker
to a nearest neighbor site can be imagined as initiating formation of an unsaturated bifunctional
monomer, while trapping would give polymers of different sizes.

This paper is organized as follows. We first present a short review
of the theory and major results of the one-dimensional QW in
Sect.~\ref{sec:qrw}. The model of CRW and QW with traps is introduced
and the survival probability is defined in Sect.~\ref{sec:model}.
In Sect.~\ref{sec:results}, the numerical simulations on the
survival probability are shown and the analytical results are given
using the correspondence to the QW with position measurement on
the line in the thermodynamic limit. Section~\ref{sec:con} is devoted to the
conclusions and the outlooks.
\section{Quantum walk}
\label{sec:qrw}
\subsection{Single-particle walk}
We consider a coined discrete-time QW on a finite linear lattice
segment with periodic boundary conditions. Denoting the total number
of sites on the lattice by $K$, the geometry is equivalent to a
ring, or a so-called $K$-cycle \cite{kendon,zn1}. In strict
mathematical terms, it is the Cayley graph of the cyclic group of
size $K$. The coin (chirality) space of a single walker is described
by $\mathcal{H}_C$ with two basis vectors $\{\ket \uparrow, \ket
\downarrow \}$. Also, the position space of a single walker on this
chain is described by $\mathcal{H}_P$ with the basis $\{\ket{k} \ :
\ k \in \mathbb{Z} / K \mathbb{Z} \}$. The Hilbert space of total
system is given by $\mathcal{H} = \mathcal{H}_C \otimes
\mathcal{H}_P$. We identify the chirality basis vectors as
\begin{equation}\label{1}
|\uparrow\rangle=\left(
\begin{array}{c}
1 \\
0 \\
\end{array}
\right), \quad |\downarrow\rangle=\left(
\begin{array}{c}
0 \\
1 \\
\end{array}
\right) \ .
\end{equation}
Each step of the particle~\footnote{Throughout this paper, we call it the step or the time. It should be noted that these have the same meaning.}
consists of a unitary coin operation
$\hat{C}$ for the chirality transformation, and a position-shift
operation $\hat{S}$. At time $t$, QW is defined by
transformation $\hat{U}^t$ with $\hat{U}$ being the unitary operator
of a walk step which is given by
\begin{equation}\label{2}
\hat{U}:=\hat{S}(\hat{C} \otimes \hat{I}),
\end{equation}
with $\hat{I}$ being the identity operator.
Throughout this paper, we assume that the coin operator $\hat{C}$ is
the Hadamard operator
\begin{equation} \label{3}
\hat{H}=\frac{1}{\sqrt{2}}\left(
\begin{array}{cc}
1 & 1 \\
1 & -1 \\
\end{array}
\right) \
\end{equation}
for simplifying the discussion. The shift position operator $\hat{S}$ is described by
\begin{equation} \label{4}
\hat{S}=|\uparrow\rangle\langle\uparrow
|\otimes\sum_{k=1}^K | k+1 \rangle\langle
k |+|\downarrow\rangle\langle\downarrow|\otimes\sum_{k=1}^K
|k-1 \rangle\langle k | \ ,
\end{equation}
with $k \in \mathbb{Z} / K \mathbb{Z}$; $K+1 \equiv 1$ and $0 \equiv K$. The wave function of
quantum walker at time $t$ can be written as
$|\psi(t)\rangle=\sum_{c,k}\psi_c(k,t)\ket{c,k}$ with
$c=\uparrow,\downarrow$. This can be rewritten as
\begin{equation}\label{5}
\ket{\psi(k,t)}=\sum_{c \in \{ \uparrow,\downarrow \}
}\psi_c(k,t)\ket{c}=\left[
\begin{array}{c}
\psi_{\uparrow}(k,t) \\
\psi_{\downarrow}(k,t) \\
\end{array}
\right],
\end{equation}
where $\psi_{\uparrow}(k,t)$ and $\psi_{\downarrow}(k,t)$
represent probability amplitudes of the particle at the site $k$
at time $t$, depending upon the internal states $\ket{\uparrow}$ and
$\ket{\downarrow}$, respectively.

At time $t$, the quantum state of the quantum walker is given by
\begin{equation}
    \ket{\psi (t)} = \hat{U}^t \ket{\psi (0)},
\end{equation}
where $\ket{\psi (0)} = \ket{\chi, m}$ is the initial state of the
coin $\ket{\chi}$ and the position $\ket{m} \ (m \in \mathbb{Z}/K\mathbb{Z})$.
Here, we define the density operator of the quantum walker as $\Phi (t) = \ket{\psi (t)}
\bra{\psi (t)}$. Then, the probability distribution of walker at position $x$ at time
$t$ can be calculated by
\begin{equation}
    P(x,t) = \sum_{c \in \{ \uparrow, \downarrow \} } \bra{c, x} \Phi(t) \ket{c, x} \ \ (x \in \mathbb{Z} / K \mathbb{Z}).
\end{equation}
This can be rewritten as
\begin{equation}\label{8}
P(x,t)=|\psi_{\uparrow}(x,t)|^{2}+|\psi_{\downarrow}(x,t)|^{2} \ .
\end{equation}
\subsection{Multi-particle walk}
One can easily envision that multi-particle random walks can be more
advantageous in search algorithms than single-particle ones. Indeed,
recent experimental progress and theoretical studies favor the
many-body random walk problem both in classical and in quantum
realms \cite{Omar1,Goyal}. As the general approach in terms of
indistinguishable, correlated and interacting particles to this
problem is too challenging to start with, we aim to comprehend the
simplest scenario in this work and consider the complications in
particular implementation settings in future studies. Let us assume
the walkers are non-interacting distinguishable particles and they
are initially uncorrelated. For $N$ such walkers, the Hilbert space
is given by a direct product of single walker spaces,
\begin{equation}\label{9}
\mathcal{H}=\bigotimes_{i=1}^{N}(\mathcal{H}_{C} \otimes \mathcal{H}_{P})_{i},
\end{equation}
with the particle label $i$. The particles walk independent of each other on the $K$-cycle so
that the time evolution of the whole system is determined by
\begin{equation}\label{10}
\hat{U}_{1,2,\ldots ,N} := \hat{U}^{\otimes N}
\end{equation}
where $\hat{U}$ is given by Eq.\,(\ref{2}) and is the same for all particles.
\begin{figure}[h!]
\begin{center}
\centering
\includegraphics[width=3.2in]{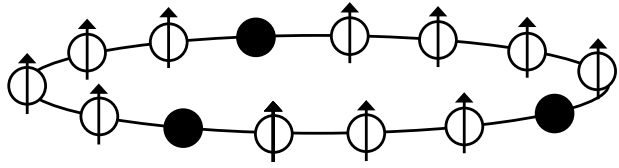}
\caption{\label{fig1} Schematic representation of multi-particle
QW on the $K$-cycle for $11$ walkers with the initial state $\ket
\uparrow$ and $3$ absorbing trapped sites (black color).}
\end{center}
\end{figure}

The initial state of $N$ walkers is given by a tensor product of the
single walker initial states as
\begin{equation}\label{11}
	\ket{\Psi (0)} = \bigotimes_{i=1}^{N} \ket{\chi,m_i}_i,
\end{equation}
where $\ket{\chi,m_i}_i$ expresses the $i$th particle state with the chirality $\ket{\chi}$
and the position $m_i = 1, \dots, K$. This is illustrated in Fig.\,\ref{fig1}.
At time $t$, the quantum state of the system becomes
\begin{equation}\label{12}
\ket{\Psi (t)}=\hat{U}_{1,2,\dots, N}^{t} \ket{\Psi (0)}.
\end{equation}

Using the reduced single-particle density matrix
\begin{equation}\label{13}
\Phi_i (t) = \Tr_{j \ne i} \ket{\Psi (t)} \bra{\Psi (t)},
\end{equation}
probability distribution of single walker at time $t$ can be
evaluated by
\begin{equation}\label{14}
P_i (x,t) = \sum_{c \in \{ \uparrow , \downarrow \} } \bra{c,x}
\Phi_i (t) \ket {c,x} \ \ (x \in \mathbb{Z} / K \mathbb{Z}).
\end{equation}
This shows that $P_i (x,t)$ can be interpreted as a conditional
probability to find a walker at site $x \in \mathbb{Z} / K \mathbb{Z}$ at time
$t$ when the particle started to walk from site $m_i$ at  $t=0$. The
complete set of $\{P_i(x,t)\}$ for all particles $i=1, \dots ,N$
can be visualized as the set of transition probabilities from
$\{ m_i \}$ to $x$ of a single particle, so that the simplest
multi-particle QW problem under study here is essentially a single
particle problem that starts to walk at a set of different initial
locations.
\section{Survival probability}
\label{sec:model}
We use the exact enumeration method for calculating the survival
probability in CRW \cite{Havlin1}, which is suitable for the
randomly distributed immobile traps on a one-dimensional lattice.
Initially, every untrapped site is occupied by a walker. At each
step, $N$ walkers perform CRW on the one-dimensional
lattice, for which the probability of finding a walker
at a particular site $P_i(x,t)$ is calculated with the sum of the
corresponding probabilities at its nearest neighbor sites divided by
two. The survival probability at time $t$ is given by
\begin{equation}\label{15}
P_r(t) = \frac{1}{N} \sum_{i = 1}^N \sum_{x = 1}^K P_{i}(x,t).
\end{equation}
Here $r$ enumerates a particular independent initial configuration
of the system. Note that we have the relation $N=K-n$ or
$N=K(1-\rho)$, where $n$ is the number of traps on the lattice and
$\rho=n/K$ is the concentration from our assumption. We take the
lattice sites at $\{x: x=1,...,K\}$. Note that we assume
non-overlapping, immobile, perfectly absorbing sites such that
$P_i(x,t)=0$ if the site $x$ is a trapping site; hence the sum is
not restricted to the untrapped sites. Furthermore, in this paper,
the initial configurations on the particles are assumed that a single particle
is only in each untrapped site.

To account for random distribution of the traps, a statistical
configurational average of mean survival probability is calculated
over different independent realizations of the initial system via
\begin{equation}\label{16}
\left\langle P\left( t \right) \right\rangle =
\frac{1}{M}\sum_{r=1}^M P_r (t),
\end{equation}
where $M$ denotes the number of different configurations.

Let us now examine the quantum analog of the survival probability
\cite{Parris2,Parris3,Jayannavar2}. In QW, quantum states of the
particles lead to non-trivial path interference effects. The quantum
states can be initialized arbitrarily before the walk starts.  At
each time step, new positions and states of quantum walkers are
determined with the unitary transformation $\hat{U}$ for each
walker.  If a state meets with an immobile trap, it gets
annihilated. In the previous section we have seen that for our
simplified case, the process is equivalent to a single-particle
problem with an ensemble of initial configurations. Thus we can use
the classical definition of the survival probability by only making
a quantum mechanical calculation of the single-particle probability
distribution.

Let us briefly consider the experimental realization of this system.
In the system of the two-dimensional ion trap experiment ~\cite{zahringer},
it seems to be possible to
manipulate the QW on the ring. Localized ion losses can be effectively
considered as the absorption traps. Also, in the system using the photon by the
waveguide, the non-linear phase gate is essentially used to realize the
QW on the circle~\cite{phd}. Combining the waveguides, the absorption traps
could to be realized.
\section{Results and discussions}
\label{sec:results}
\subsection{Survival probability in QW on a finite one-dimensional trapped lattice}
We numerically analyze the dynamics of the survival probability in QW for three
different initializations of the system. In the first case, all
quantum states at the untrapped sites are initialized as
$\ket{\uparrow}$. In the second case, the initial states are randomly
assigned either $\ket{\uparrow}$ or $\ket{\downarrow}$. In the last
case, all states are chosen as superpositions
$\frac{1}{\sqrt{2}}(\ket{\uparrow} + i\ket{\downarrow})$. We shall
respectively call them as up, mixed and symmetric initializations.
\begin{figure}[!]
\begin{center}
\subfigure[\hspace{0.01cm}]{\label{fig:sub:2a}
\includegraphics[width=2.2in]{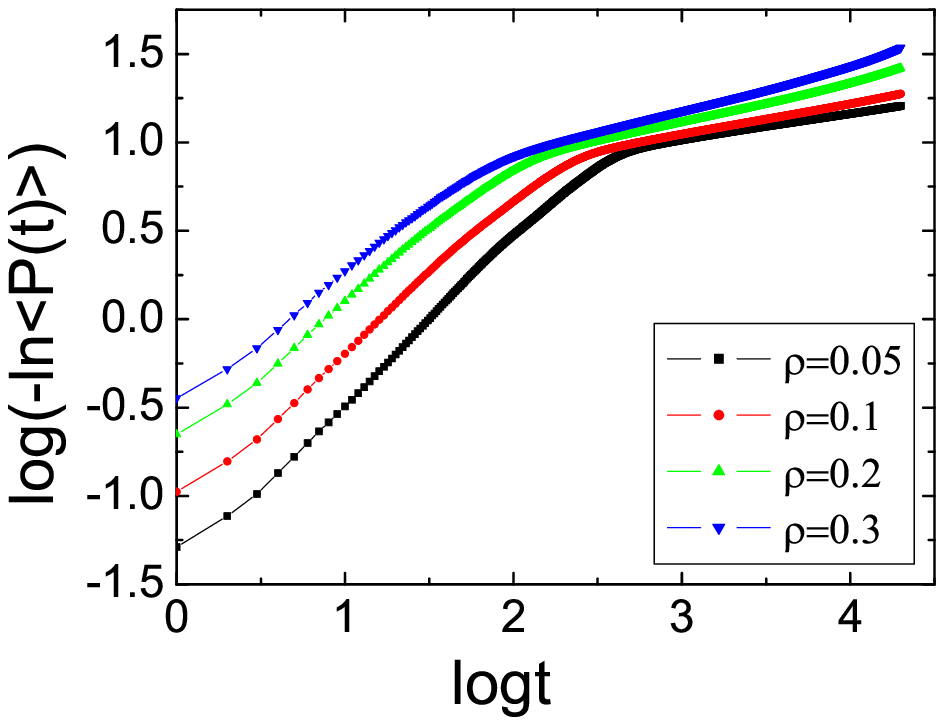}}
\hspace{0.01cm} \subfigure[\hspace{0.01cm}]{\label{fig:sub:2b}
\includegraphics[width=2.2in]{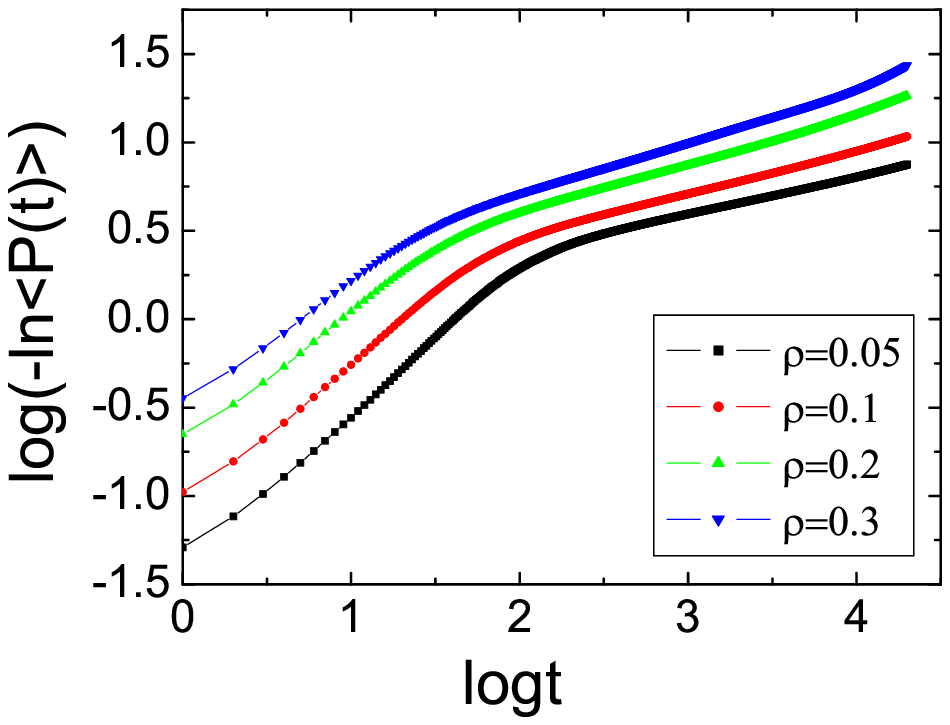}}
\hspace{0.01cm} \subfigure[\hspace{0.01cm}]{\label{fig:sub:2c}
\includegraphics[width=2.2in]{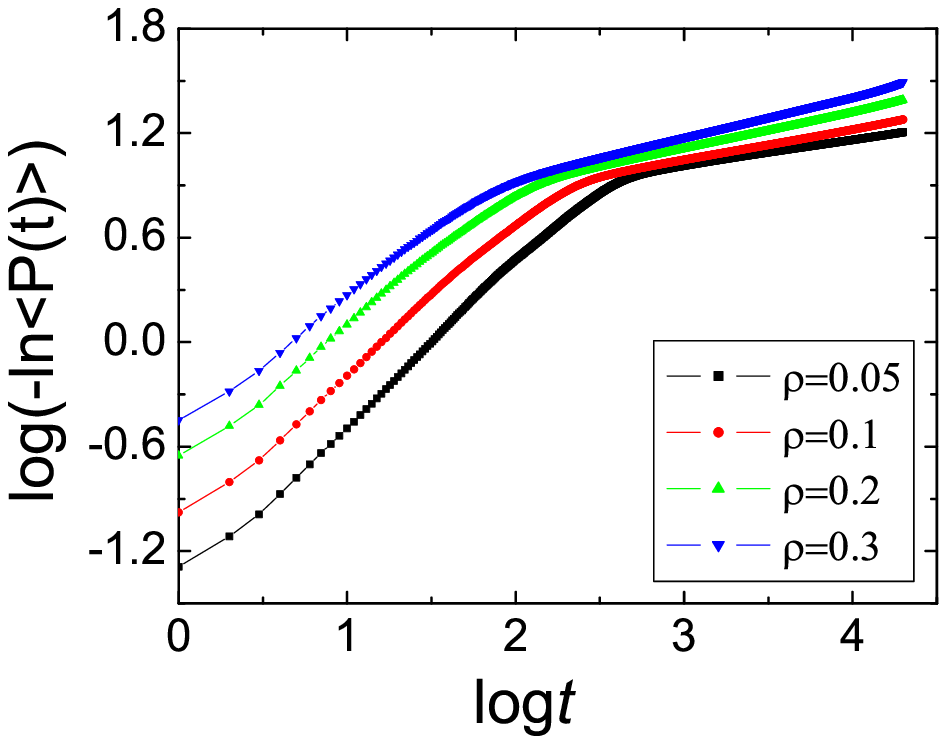}}
\caption{\label{fig2}(a) The time dependence of the survival
probability in the QW on a lattice of $K=101$ sites for $T=20000$,
$M=10000$, $\rho=0.05$, $\rho=0.1$, $\rho=0.2$ and $\rho=0.3$ with
the initializations a) $\ket{\uparrow}$ (up), (b) randomly distributed
$\ket{\uparrow}$ or $\ket{\downarrow}$ (mixed), and (c)
$\frac{1}{\sqrt{2}}(\ket{\uparrow}+i\ket{\downarrow}) $(symmetric).}
\end{center}
\end{figure}

Typical simulation results are reported in
Figs.\,\ref{fig2}(a)--(c), for the three cases of initialization of
the QW, and for different trap densities, $\rho=0.05$, $\rho=0.1$,
$\rho=0.2$, and $\rho=0.3$. The figures are plotted in a convenient
double logarithmic scale. While the survival probability exhibits
the expected behaviors of decrease in time and being less at higher
trap densities, it invites a closer look due to some non-trivial
qualitative changes in its dynamics. All the different
initializations lead to two qualitatively different dynamical
regimes of survival probability. These two regimes makes a crossover
at a certain time point, $t_{c}$, whose location depends on the trap
density. The crossover time $t_c$ appears at $t_{c}\approx 25/\rho$
for up and symmetric initializations, and at $t_{c}\approx 8/\rho$
for mixed initialization. Figure\,\ref{fig2} is plotted for $K=101$
sites but we also tried different lattice sizes and found similar
results to Fig.\,\ref{fig2}.

\begin{figure}[h!]
\begin{center}
\centering
\includegraphics[width=2.7in]{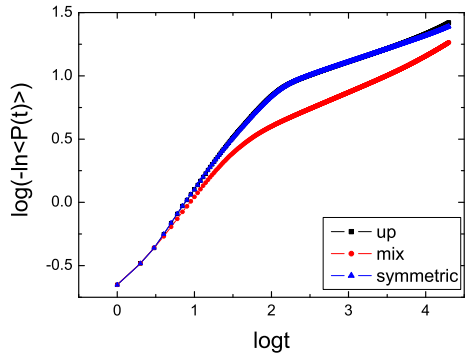}
\caption{\label{fig3} The time dependence of the survival
probability in the QW on a lattice of $K=101$ sites for $T=20000$,
$M=10000$, $\rho=0.2$ with the initializations $\ket{\uparrow}$ (up), randomly distributed
$\ket{\uparrow}$ or $\ket{\downarrow}$ (mixed) and $\frac{1}{\sqrt{2}}(\ket{\uparrow}+i\ket{\downarrow})$ (symmetric).}
\end{center}
\end{figure}

As seen in Fig.\,\ref{fig3}, the ``mixed" initial configuration behaves dynamically different than
``up" and ``symmetric" cases. This is due to the profound quantum character of pure states
in contrast to the statistical mixture,  evolving more closer to classical walk.
Pure quantum states benefit the fast spread of the quantum walk in the Rosenstock regime more than
a statistical mixture can do. Classical walks perform worst
in this regime as we shall argue more below.

\begin{figure}[!]
\begin{center}
\includegraphics[width=2.7in]{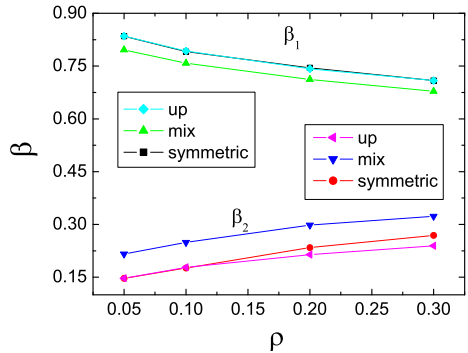}
\caption{\label{fig4} Dependence of the decay parameters
$\beta_{1,2}$ on the low trap density $\rho$ in a lattice of $K=101$
sites, in cases of the up, mixed and symmetric initializations.}
\end{center}
\end{figure}

Before and after the crossover point, the survival probability
exhibits a linear dependence on time in the double logarithmic
scale. By increasing $M$ the linear behavior becomes more evident
including the end and the beginning times.
After that it is simple
to make linear fits to the curves in the regimes from $t=1$ to $t=t_c$
and from $t=t_c$ to $t=T$, the end of the walk. The line fits yield the
Kohlrausch-Williams-Watts stretched exponential function
\cite{Phillips} description of the survival
probabilities, given by
\begin{equation} \label{17}
\left\langle P \left( t\right) \right\rangle \sim \exp \left[ -
t^{\beta} \right],
\end{equation}
where the stretching exponent
$0<\beta<1$ determines the decay rate of $\left\langle P \left(
t \right)\right\rangle$. It gets different values, $\beta_1$ and
$\beta_2$, before and after $t_c$, respectively. Their dependence on trap
density $\rho$ is shown in Fig.\,\ref{fig4}. $\beta_{1}$
decreases monotonically with the increasing $\rho$, whereas
$\beta_{2}$ increases with it.

The value of $\beta_1$ at low $\rho$ can be understood following the
classical RS approximation method. When $s_t$ is the number of
distinct sites visited at time $t$, the probability of being not
absorbed for a single walker can be written as $p_t=(1 -
\rho)^{s_t}$. Formally the mean probability can be expressed in the
form $P(t)= \langle p_t \rangle= \langle \exp{(-\lambda s_t)}
\rangle$ with $\lambda= - \ln{(1-\rho)}$. Employing the RS
approximation for short time and small $\rho$, we get $P(t) \approx
\exp{(-\lambda \langle s_t \rangle)}$. For the CRW, $\langle
s_t\rangle\sim\sqrt{t}$ gives the usual RS scaling form. For the QW,
however, the ballistic spread up of the quantum walkers allows for
$\langle s_t \rangle \sim t$~\cite{Konno3,Konno4} so that the
equivalent rate of survival is enhanced to $\beta_1 \sim 1$ in the
quantum analog of the RS regime.

The maximal value of $\langle s_t \rangle$ is associated with the
screening or penetration length that measures the distance between
the starting and the trapping site of the walkers. Quantum walkers
have larger penetration lengths than classical ones and as such can
survive at asymptotic times provided that they start in larger
clusters. The probability of such configurations are exponentially rare
with the size of the segment while the quantum spreading is a power
law (quadratic) gain relative to classical walk. So it is necessary
to find indeed large voids (relatively larger than their classical
counterparts) to ensure quantum walkers can survive. In the quantum analog of the DV
regime, the size of the dominating, remaining clusters with quantum
walkers, therefore would be larger than the classical DV regime.
Averaging over such slowly decaying large voids would then lead to
the survival probability decaying slower than the DV regime of
classical diffusion. Similar observation for the case of continuous-time
quantum transport gives $-\ln{\langle P(t) \rangle}= t^{1/4}$
\cite{Parris2}. In our case, $\beta_2$ is close to this value up to
$\rho<0.3$ as demonstrated in Fig. \ref{fig4}.

The Flory-type heuristic arguments for $\beta_2$ and RS
approximation for $\beta_1$ justify that the numerically observed
crossover in Fig. \ref{fig2} is indeed the quantum analog of
classical RS-to-DV transition. It is known that such a crossover can
happen only at long times and hard to observe in the CRW. In classical
systems, $t_c$ can be reduced either by increasing $\rho$, which is
especially efficient for one dimension \cite{schotland}, or by using
systems with large diffusive constants \cite{schotland}.
Remarkably, decrease of $t_c$ with $\rho$ is also observed in Fig.
\ref{fig2} for QW. Increase of $\rho$ however would mean to loose
the benefits of the quantum coherence in QW due to
quantum-to-classical transition that happens at high $\rho$
\cite{Gonulol,romanelli,chandra2007}. On the other hand, strong
de-localization of quantum walkers contributes significantly for further reduction of
$t_c$ in the QW.
As such, we expect that the quantum analog of RS-to-DV crossover can
occur earlier than CRW. As the decay of survival probability of QW
is even slower than classical diffusion in quantum DV regime, this
makes the quantum RS-to-DV crossover a serious limitation to
implement QW-based quantum search algorithms even for relatively
small number of traps (or targets). The usual limitation factor of
quantum-to-classical transition is an issue only for high $\rho$. In
the next section, we shall verify our predictions and also
investigate the effect of large trap concentrations. Furthermore, we
will analytically show the crossover in the thermodynamic limit.
\subsection{Survival probability in CRW vs QW}
\begin{figure}[!]
\begin{center}
\subfigure[\hspace{0.01cm}]{\label{fig:sub:5a}
\includegraphics[width=2.5in]{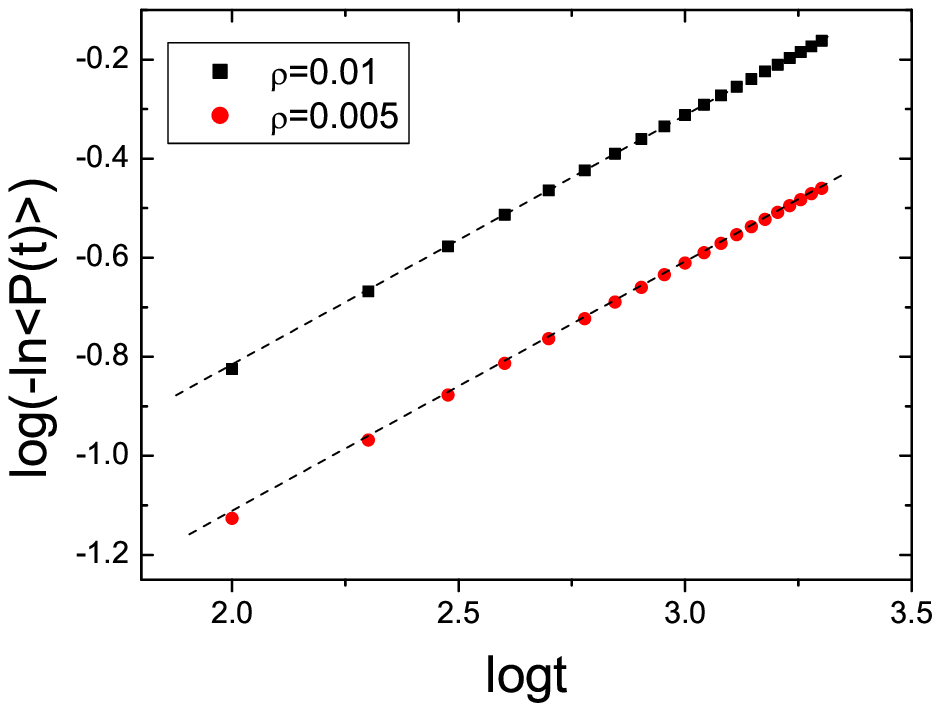}}
\subfigure[\hspace{0.01cm}]{\label{fig:sub:5b}
\includegraphics[width=2.5in]{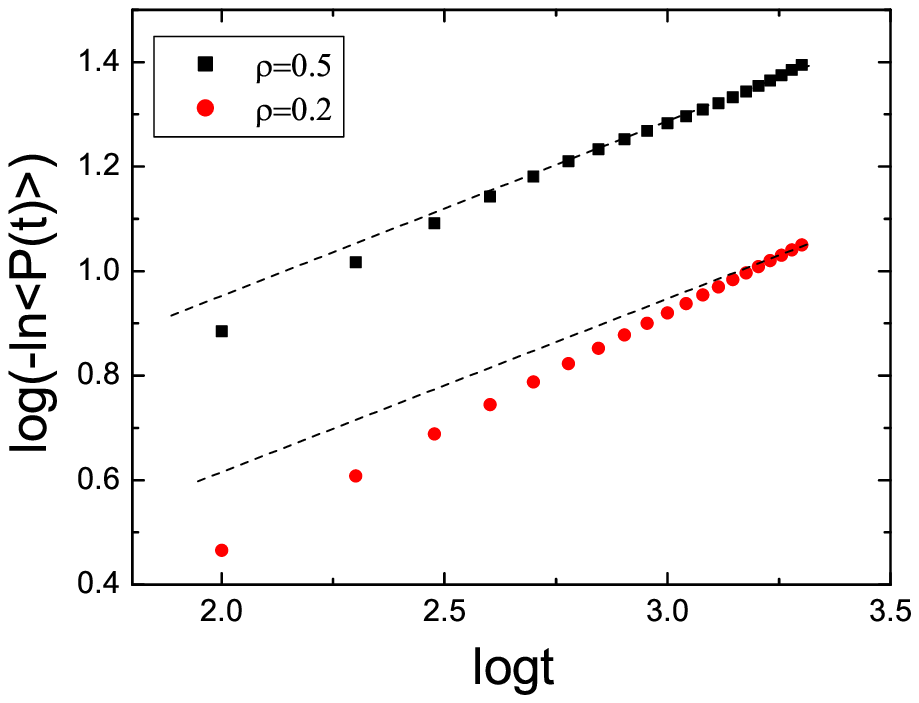}}
\caption{\label{fig5} Time dependence of the survival probability in
CRW on a lattice of $K = 50000$ sites for (a)
 $T = 2000$, $M = 100$ and $\rho=0.01$, $\rho=0.005$, where the
broken lines represent the slope of $\beta=1/2$, (b)
 $T = 2000$, $M = 100$ and $\rho=0.2$, $\rho=0.5$ where the
broken lines represent the slope of $\beta=1/3$.}
\end{center}
\end{figure}

Short and long time behaviors of the survival probability in the CRW
and evidence of RS-to-DV crossover are shown in Fig.\,\ref{fig5}. In
short time behavior, the survival probability at low trap densities
comply with RS behavior and fit well to $\beta=1/2$ in
Fig.\,\ref{fig:sub:5a}. If the trap density is increased, the slope
decreases and approaches $1/3$ for large values of $\rho$. DV
behavior emerges in Fig.\,\ref{fig:sub:5b}. These analytical values
are strictly valid for thermodynamically large system. Convergence
to the asymptotic DV scaling form is faster in case of higher trap
concentrations.

\begin{figure}[!]
\begin{center}
\subfigure[\hspace{0.01cm}]{\label{fig:sub:6a}
\includegraphics[width=2.5in]{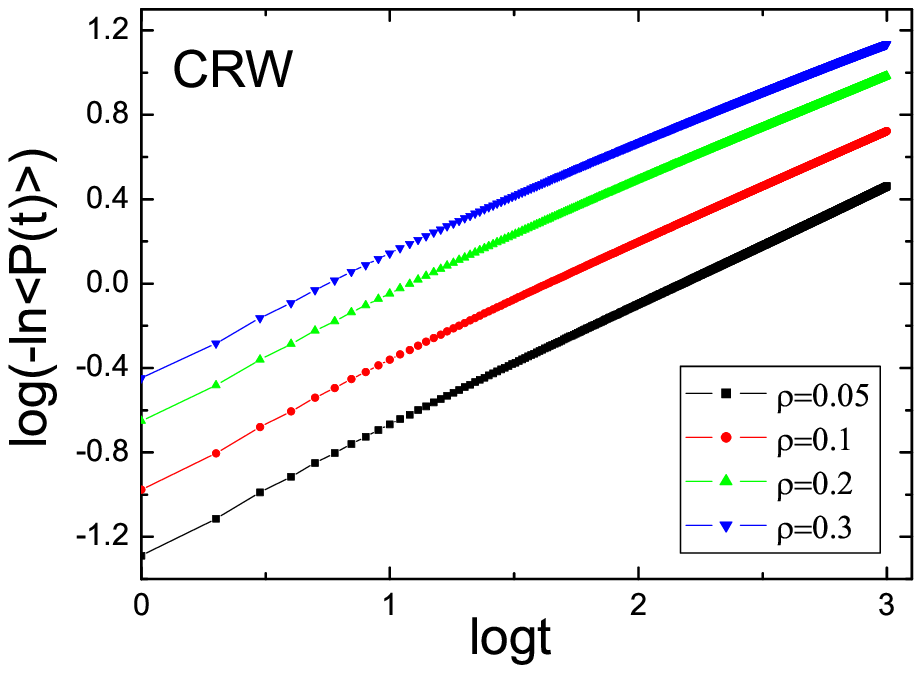}}
\subfigure[\hspace{0.01cm}]{\label{fig:sub:6b}
\includegraphics[width=2.5in]{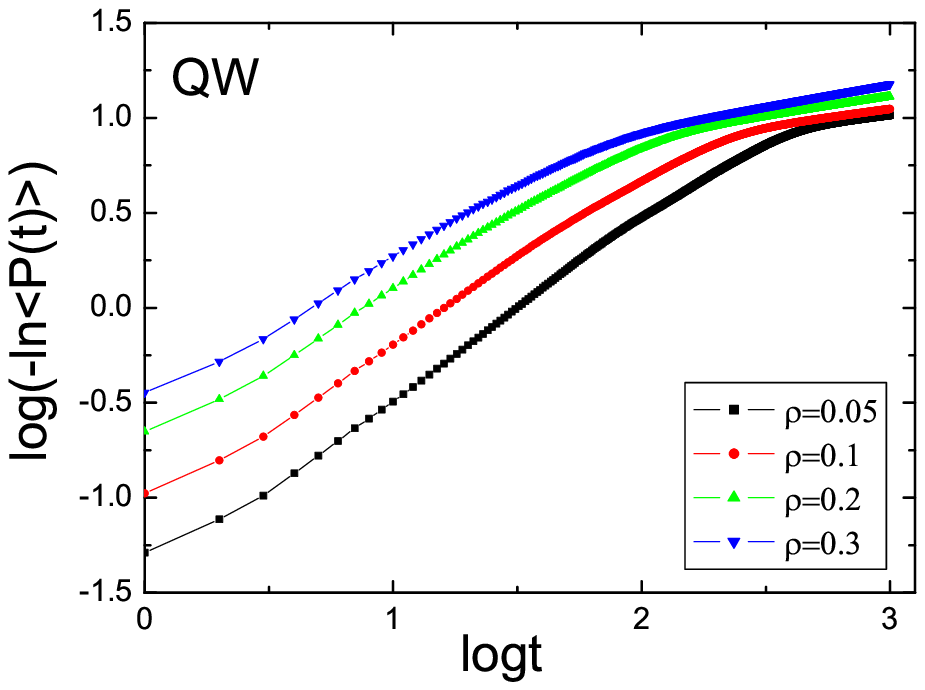}}
\caption{\label{fig6} The time dependence of the survival
probability for a lattice of $K=101$ sites, $T=1000$, $M=100000$
(a) in the CRW, (b) in the QW with the initial state
$\ket{\uparrow}$.}
\end{center}
\end{figure}
\begin{figure}[!]
\begin{center}
\includegraphics[width=3in]{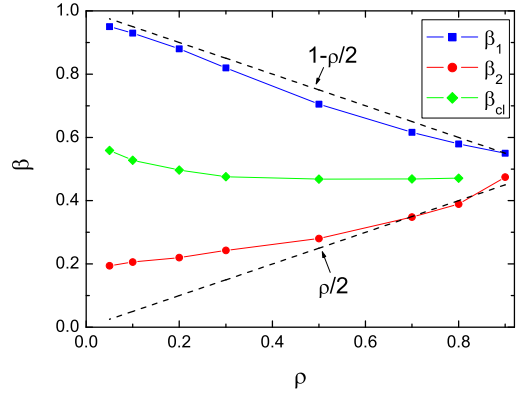}
 \caption{\label{fig7}Dependence of the decay parameter $\beta$ on
the trap density $\rho$ for a lattice of $K = 101$ sites and $t=1000$ time step in the CRW and
the QW with initial state $\ket{\uparrow}$. Dashed lines represent analytical fitting results.}
\end{center}
\end{figure}
To compare the CRW and QW, we consider a lattice of
$K=101$ sites and take $T=1000$. The time dependence of survival
probability is shown in Fig.\,\ref{fig:sub:6a} for the CRW and
Fig.\,\ref{fig:sub:6b} for the QW. We choose the initialization that
gives the longest $t_c$, to consider the worst situation for the QW.
Even for this case, we see that quantum RS-to-DV crossover happens
while the CRW is still in the classical RS regime. In particular for
low $\rho$, highly distinct and clear crossover can be observed in
the QW.

From the plots on survival probability in Fig.\,\ref{fig6}, the
influence of $\rho$ on the scaling forms can be systematically
investigated by Fig.\,\ref{fig7}, in which the stretching exponents
are plotted as functions of trap density $\rho$. As the trap
concentration increases, $\beta_1$ and $\beta_{cl}$ decrease. In
contrast, $\beta_2$ increases with $\rho$. As noted earlier, the
sharpest transition between quantum RS and DV regimes happens at low
$\rho$. In high trap densities, their separation shrinks and both
$\beta_{1,2}$ converges to the classical exponent $\beta_{cl}$. This
is in accordance with the expectation that for such high $\rho$,
decoherence transition of QW to CRW should occur. Figure\,\ref{fig7}
gives strong and clear evidence that the dynamical transition in
Fig.\,\ref{fig:sub:6b}, between early and longer time scaling forms
of QW, is not a quantum-to-classical transition, but the true
quantum analog of classical RS-to-DV crossover.

\begin{figure}[!]
\begin{center}
\includegraphics[width=3in]{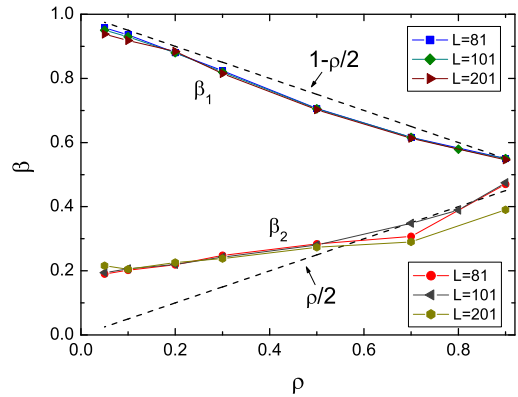}
 \caption{\label{fig8}Dependence of the decay parameter $\beta$ on
the trap density $\rho$ for lattice of $K = 81$, $K=101$ and $K=201$ sites, $t=1000$ time step in
the QW with initial state $\ket{\uparrow}$.}
\end{center}
\end{figure}

\subsection{Single-particle QW with position measurement and survival probability in multi-particle QW}
In this subsection, we analytically show the relationship between
the survival probability and the QW with position measurement on the line.

Let us recapitulate the QW with position measurement on the
line~\cite{CKSS1,CKSS2,CKSS3}. Here, we replace the position
Hilbert space to $\tilde{\mathcal{H}}_P = \{ \ket{z} : z \in \mathbb{Z} \}$.
The one-step dynamics is given by
\begin{eqnarray}
	\Phi (t + 1) & = (1 - p) U \Phi (t) U^{\dagger} \nonumber\\
	& + p \ket{\chi} \bra{\chi} \otimes \left[ \Tr_C \sum_{z, z^{\prime} \in \mathbb{Z}} \left[ ( \hat{I} \otimes \ket{z}
	\bra{z} ) \hat{U} \Phi (t) \hat{U}^{\dagger} ( \hat{I} \otimes \ket{z^{\prime}} \bra{z^{\prime}} ) \right] \right], \label{18}
\end{eqnarray}
where $\Phi (0) = \ket{\tilde{\psi}(0)} \bra{\tilde{\psi}(0)}$ with
$\ket{\tilde{\psi}(0)} = \ket{\chi, 0}$ and $p \in [0, 1]$. This model can be
taken as the position measurement of the one-dimensional QW with
probability $p$. When $p = 1/t^{\gamma} \ (0 \leq \gamma \leq 1)$,
the asymptotic behavior of the QW with position measurement is $\langle
s^{(D)}_t \rangle \sim t^{(1 + \gamma)/2}$~\cite{CKSS1, CKSS2}.

In the case of the thermodynamic limit, $K \to \infty$ with fixed
$\rho$, and the sufficiently large $t$, many-particle QW on the
$K$-cycle can be reduced to the single-particle QW on the line as
follows. For uncorrelated quantum walkers and quantum coin in our
model of multi-particle QW, the event of annihilation of the walker
reaching a trap site is equivalent to a position measurement.
The mean probability that the
particle reaches the trapped site at time $t$ is $p = t^\rho / t =
(1/t)^{(1-\rho)}$. Therefore, it is possible to apply the result on
the QW with position measurement on the line to this system to obtain the
asymptotic behavior of the QW to arrive at the trapped site as
$\langle s^{(T)}_t \rangle \sim t^{1-\frac{\rho}{2}}$. This can be
taken the mean free path. In the thermodynamic limit, we can apply
the central limit theorem to obtain that the survival probability is
the exponential decay for the mean free path as
\begin{equation} \label{19}
\langle P (t) \rangle \sim \exp \left[ - \frac{\langle s^{(NT)}_t
\rangle}{\langle s^{(T)}_t \rangle} \right] \sim \exp \left[ -
t^{\frac{\rho}{2}} \right],
\end{equation}
where $\langle s^{(NT)}_t \rangle \sim t$ is the mean free path
without the trap site, {\it i.e.}, the QW behavior without position
measurement.

Let us reconsider the thermodynamic limit with the fixed $\rho$ for
the two types: $t \ll K$ and $t \sim K$. In the first case, {\it
i.e.}, before the crossover time, the survival probability can be
approximately taken as the small $t$. That is, it is impossible to
directly apply Eq. (\ref{19}). Analogous to the discussion in the
above section, the survival probability can be rewritten as
\begin{equation} \label{20}
\langle P (t) \rangle \approx 1 - \langle s^{(T)}_t \rangle  \sim 1
- t^{\left( 1 - \frac{\rho}{2} \right)}
\end{equation}
From $\langle P (t) \rangle \sim \exp \left[ -t^{\beta_1}\right]
\approx 1 - t^{\beta_1}$, we obtain
\begin{equation} \label{21}
\beta_1 = 1 - \frac{\rho}{2}
\end{equation}
In the second case,{\it i.e.}, after the crossover time, on the
other hand, Eq. (\ref{19}) can express the exponential decay.
Therefore, we directly obtain
\begin{equation} \label{22}
\beta_2 = \frac{\rho}{2}
\end{equation}
These analytical results can be compared with the numerical results of
Fig.\,\ref{fig7}. While these show good agreement, we cannot see
the finite-size effects as seen in Fig.\,\ref{fig8}. Also, our numerical
calculation is used in the same and specific chirality state for the
multi-particle QW and the Hadamard coin. When we remove these
conditions, our analytical observation is unchanged in the
thermodynamic limit since the essential part of the proof the limit
distribution is only the identical distribution for the single
particle~\cite{CKSS2}.
\section{Conclusion and Outlook}
\label{sec:con}
We have investigated the time
dependence of the survival probability in discrete Hadamard QW on a
K-cycle with random, static, perfect traps.
We have found that
the survival probability exhibits a piecewise stretched exponential
character. In the early time regime, it decays faster than that of
CRW, while in the late time regime it decays slower. The crossover
time between two regimes decreases with trap density $\rho$. By analytical and
heuristic arguments, we have identified the dynamical transition
between two regimes as quantum analog of the RS-to-DV crossover in
classical diffusion. We have shown that quantum RS-to-DV crossover
can happen earlier than its classical counterpart.
At high trap concentrations quantum-to-classical transition
happens. At low trap concentrations, even if quantum-to-classical
transition does not play a role, quantum RS-to-DV crossover has
found to be a serious limitation on the benefits of quantum
coherence, such as quadratic speeding up in implementations of QW-based
quantum search algorithms.

As an outlook of present work, consideration of larger dimensional systems with
probabilistic or state-dependent traps and interacting walkers could make the
results more suitable for applications and experimental realizations~\cite{mayer}.
From a more fundamental point of view, further investigations
of the quantum RS-DV scaling transitions can be performed for in terms of quantum Zeno effect in
QW~\cite{Chand} or in relation to random quenched disorder~\cite{Schreiber}. A more direct and rigorous
generalization of Flory mean field theory to the trapped quantum random walk can also be pursued.
An extension of our work for the question of quantum RS-to-DV transition in the case of continuous
QW would be of interest in the light of the recent realizations of multi-agent continuous QW \cite{peruzzo}.

\ack
The authors thank Zafer Gedik,
Hamza Polat, Ismail Hakki Duru, Kaan Guven and Jun-ichi Inoue for
many discussions and helpful comments. Two of the authors
(EA and \"{O}EM) acknowledge warm hospitality and support by
the Dokuz Eyl\"{u}l University. This work was supported by the
TUBITAK (The Scientific and Technological Research Council of Turkey)
under research project (No.109T681), \.{I}stanbul University (No.3660 and No.6942),
and JSPS Research Fellowships for Young Scientists (No.21008624). OEM gratefully acknowledges the
support by the DPT (T.C. Prime Ministry State Planning Organization) under the
project of National Quantum Cryptology Center for UEKAE (National
Research Institute of Electronics and Cryptology). YS is supported by
Global Center of Excellence Program ``Nanoscience and Quantum Physics" at Tokyo Institute of Technology.


\section*{References}
\begin{thebibliography}{99}
\bibitem{barber} Barber M N and Ninham B W 1970 {\it Random and Restricted Walks: Theory and Applications} (New York: Gordon and Breach)

\bibitem{ambainis2001} Ambainis A, Bach E, Nayak A, Vishwanath A and Watrous J 2001 {\it Proceedings of the 33rd Annual ACM Symposium on the Theory of Computing} (New York: ACM) p 37

\bibitem{kempe} Kempe J 2003 {\it Contemp. Phys.} {\bf 44} 307

\bibitem{shenvi} Shenvi N, Kempe J and Whaley K Birgitta 2003 {\it Phys. Rev. A} {\bf 67} 052307

\bibitem{zahringer} Z\"{a}hringer F, Kirchmair G, Gerritsma R, Solano E, Blatt R and Roos C F 2010 {\it Phys. Rev. Lett.} {\bf 104} 100503

\bibitem{schmitz} Schmitz H, Matjeschk R, Schneider Ch, Glueckert J, Enderlein M, Huber T and Schaetz T 2009 {\it Phys. Rev. Lett.} {\bf 103} 090504

\bibitem{schreiber} Schreiber A, Cassemiro K N,  Poto\v{c}ek V, G\'{a}bris A, Mosley P J, Andersson E, Jex I and Silberhorn Ch 2010 {\it Phys. Rev. Lett.} {\bf 104} 050502

\bibitem{lu} Lu D, Zhu J, Zou P, Peng X, Yu Y, Zhang S, Chen Q and Du J 2010 {\it Phys. Rev. A} {\bf 81} 022308

\bibitem{broome} Broome M A, Fedrizzi A, Lanyon B P, Kassal I, Aspuru-Guzik A and White A G 2010 {\it Phys. Rev. Lett.} {\bf 104} 153602

\bibitem{Konno2} Konno N 2009 {\it Stochastic Models} {\bf 25} 28

\bibitem{Gonulol} Gonulol M, Aydiner E and Mustecaplioglu O E 2009 {\it Phys. Rev. A} {\bf 80} 022336

\bibitem{kendon2006} Kendon V 2007 {\it Math. Struct. in Comp. Science}  {\bf 17} 1169

\bibitem{CKSS1} Shikano Y, Chisaki K, Segawa E and Konno N 2010 {\it Phys. Rev. A} {\bf 81} 062129

\bibitem{romanelli} Romanelli A, Siri R, Abal G, Auyuanet A and Donangelo R 2005 {\it Physica A} {\bf 347} 137

\bibitem{chandra2007} Chandrashekar C M, Srikanth R and Banerjee S 2007 {\it Phys. Rev. A} {\bf 76} 022316

\bibitem{kendon} Kendon V and Tregenna B 2003 {\it Phys. Rev. A} {\bf 67} 042315.

\bibitem{Whitfield} Whitfield J D, Rodr\'{i}guez-Rosario C A and Aspuru-Guzik A 2010 {\it Phys. Rev. A} {\bf 81} 022323

\bibitem{havlin_review} Havlin S and Ben-Avraham D 1987 {\it Adv. Phys.} {\bf 36} 696

\bibitem{trapCRW1} Weiss G H 1999 {\it Aspects and Applications of the Random Walk} (Amsterdam: North-Holland)

\bibitem{trapCRW4} Gallos L K and P Argyrakis 2001 {\it Phys. Rev. E} {\bf 64} 051111

\bibitem{anlauf} Anlauf J K 1984 {\it Phys. Rev. Lett.} {\bf 52} 1845

\bibitem{RS} Rosenstock H B 1970 {\it J. Math. Phys.} {\bf 11} 487

\bibitem{Barkema} Barkema G T, Biswas P and van Beijeren H 2001 {\it Phys. Rev. Lett.} {\bf 87} 170601

\bibitem{DV} Donsker M and Varadhan S R S 1979 {\it Commun. Pure Appl. Math.} {\bf 32} 721

\bibitem{Havlin1} Havlin S, Weiss G H, Kiefer J E and Dishon M 1984 {\it J. Phys. A: Math. Gen.} {\bf 17} L347

\bibitem{Grassberger} Grassberger P and Procaccia I 1982 {\it J. Chem. Phys.} {\bf 77} 6281

\bibitem{haus} Haus J W and Kehr K W 1987 {\it Phys. Rep.}{\bf 150} 263

\bibitem{lifshitz} Lifshitz I M 1964 {\it Adv. Phys.} {\bf 13} 483

\bibitem{Jayannavar3} Jayannavar A M and K\"{o}hler J 1990 {\it Phys. Rev. A} {\bf 41} 3391

\bibitem{nieuwenhuizen} Nieuwenhuizen Th M 1989 {\it Phys. Rev. Lett.} {\bf 62} 357

\bibitem{schotland} Schotland J 1988 {\it J. Chem. Phys.} {\bf 88} 907

\bibitem{Parris2} Parris P E 1989 {\it Phys. Rev. B} {\bf 40} 4928

\bibitem{mulken} M\"{u}lken O,  Blumen A, Amthor T, Giese C, Reetz-Lamour M and Weidem\"{u}ler M 2007 {\it Phys. Rev. Lett.} {\bf 99} 090601

\bibitem{agliari} Agliari E, M\"{u}lken O and Blumen A 2010 {\it Int. J. Bifurcation Chaos} {\bf 20} 271

\bibitem{florry} Flory P J 1971 {\it Principles of Polymer Chemistry} (Ithaca: Cornell University)

\bibitem{zn1} Aharonov D, Ambainis A, Kempe J and Vazirani U 2001 {\it Proceedings of the 33rd ACM Symposium on Theory of Computing} (New York: ACM) p 50

\bibitem{Omar1} Omar Y, Paunkovic N, Sheridan L and Bose S 2006 {\it Phys. Rev. A} {\bf 74} 042304

\bibitem{Goyal} Goyal S K and Chandrashekar C M 2010 {\it J. Phys. A: Math. Theor.} {\bf 43} 235303

\bibitem{Parris3} Parris P E 1989 {\it Phys. Rev. Lett.} {\bf 62} 1392

\bibitem{Phillips} Phillips J C 1996 {\it Rep. Prog. Phys.} {\bf 59} 1133

\bibitem{Jayannavar2} Jayannavar A M 1991 {\it Solid State Comm.} {\bf 77} 457

\bibitem{phd} Linjordet T 2009 Ph. D dissertation at Department of Physics and Engineering, Macquarie University, arXiv:1010.3784

\bibitem{Konno3} Konno N 2002 {\it Quantum Inf. Proc.} {\bf 1} 345

\bibitem{Konno4} Konno N 2005 {\it J. Math. Soc. Jpn.} {\bf 57} 1179

\bibitem{CKSS2} Chisaki K, Konno N, Segawa E and Shikano Y 2010 arXiv:1009.2131

\bibitem{CKSS3} Shikano Y 2010 to be published in {\it AIP Conf. Proc. as the proceedings of Advances in Quantum Theory} ed Khrennikov A

\bibitem{mayer} Mayer K, Tichy M C, Mintert F, Konrad T, and Buchleitner A 2010 arXiv:1009.5241

\bibitem{Chand} Chandrashekar C M 2010 {\it Phys. Rev. A} {\bf 82} 052108

\bibitem{Schreiber} Schreiber A, Cassemiro K N, Poto\v{c}ek V, G\'{a}bris A, Jex I, and Silberhorn Ch 2011 arXiv:1101.2638

\bibitem{peruzzo} Peruzzo A, Lobino M, Matthews J C F, Matsuda N, Politi A, Poulios K, Zhou X, Lahini Y, Ismail N, W\"{o}rhoff K, Bromberg
Y, Silberberg Y, Thompson M G and O'Brien J L 2010 {\it Science} {\bf 329} 1500
\end{thebibliography}
\end{document}